\documentclass[%
 reprint,
superscriptaddress,
showpacs,preprintnumbers,
 amsmath,amssymb,
 aip,
]{revtex4-1}

\usepackage{graphicx}
\usepackage{dcolumn}
\usepackage[version=4]{mhchem}
\usepackage{xcolor}

\begin{document}
\title{
Spectroscopic coincidence experiments in transmission electron microscopy
}

\author{Daen Jannis}
\altaffiliation[]{These authors contributed equally to the work.}
\affiliation{EMAT, University of Antwerp, Groenenborgerlaan 171, 2020 Antwerpen, Belgium} 
\author{Knut M\"uller-Caspary}
\altaffiliation[]{These authors contributed equally to the work.}
\affiliation{EMAT, University of Antwerp, Groenenborgerlaan 171, 2020 Antwerpen, Belgium}
\affiliation{Ernst Ruska-Centre for Microscopy and Spectroscopy with Electrons and Peter Gr\"unberg Institute,
	Forschungszentrum J\"ulich, 52425 J\"ulich, Germany}
\author{Armand B\'{e}ch\'{e}}
\affiliation{EMAT, University of Antwerp, Groenenborgerlaan 171, 2020 Antwerpen, Belgium}
\author{Andreas Oelsner}
\affiliation{Surface Concept GmbH, Am S\"agewerk 23a, 55124 Mainz, Germany}
\author{Johan Verbeeck}
\affiliation{EMAT, University of Antwerp, Groenenborgerlaan 171, 2020 Antwerpen, Belgium}

\begin{abstract}
	We demonstrate the feasibility of coincidence measurements in a conventional transmission electron microscope, revealing the temporal correlation between electron energy loss spectroscopy (EELS) and energy dispersive X-ray (EDX) spectroscopy events. We make use of a delay line detector with picosecond time resolution attached to a  modified EELS spectrometer. We demonstrate that coincidence between both events, related to the excitation and de-excitation of atoms in a crystal, provides added information not present in the individual EELS or EDX spectra. In particular, the method provides EELS with a significantly suppressed or even removed background, overcoming the many difficulties with conventional parametric background fitting as it uses no assumptions on the shape of the background, requires no user input and does not suffer from counting noise originating from the background signal. This is highly attractive, especially when low concentrations of elements need to be detected in a matrix of other elements. 


\end{abstract}
\maketitle
When a high-energy electron interacts with a material, a large variety of scattering processes can occur. For inelastic scattering, the electron energy loss is typically subdivided into two regions. The low-loss region, ranging approximately from 0 to 50 eV, is dominated by plasmon, interband and excitonic excitations. The high energy-loss domain, is dominated by inner-shell excitations where electrons, mainly in the K,L and M shells, are excited to higher unoccupied states in the crystal. After this process, the atom de-excites via X-ray or Auger electron emission. Since the atomic energy levels are characteristic for the type of atom, local chemical information can be obtained by measuring the energy of the secondary particles (X-rays and Auger electrons) or the energy of the outgoing high-energy electrons. This is used routinely and with up to atomic resolution in transmission electron microscopy (TEM) in energy dispersive X-ray spectrometry (EDX, detecting X-rays) \cite{VanCappellen1994, Williams2009,Schlossmacher2010} and electron-energy loss spectroscopy (EELS, detecting the energy loss of accelerated electron)\cite{Egerton2011, Tan2011, Tan2012, Turner2012, Muller1999, Dudarev1998}.\par 
In modern TEM instruments, both methods can be applied in parallel, scanning a sub-angstrom  probe over the sample and collecting both EELS and EDX spectra simultaneously.~Algorithms are then applied to correlate this multi-detector signal, sometimes together with e.g. the high angle annular dark field (HAADF) signal, into a meaningful representation of the details of the sample \cite{Ali1999,SPIEGELBERG2017}.~As much as this field is in development, an important factor is missed in this process: the excitation (EELS) and de-excitation (EDX) are intrinsically coupled and are expected to take place in very close temporal succession. The idea of observing these temporal correlations has been applied by Kruit et al. making use of real time event filtering with mixed digital/analog electronics \cite{Kruit1984}, imposing a predetermined energy window on the EELS or EDX. We expand on this idea with an updated setup that is capable of digitally storing all detected events together with their time of occurrence. This setup keeps all detected events and allows for much more extensive post processing while using a detector setup that is less complicated profiting from advances in high speed digital electronics since the 1980-ies. Since EELS and EDX , both probe the same process, one could infer that no extra information would be gained by measuring the temporal correlation between the two signals. However, there are clear distinctions between both methods that are partly technological and partly physical. A technological limitation is the limited energy resolution of current EDX detectors, hiding fine details that would provide bonding information. EELS suffers from the presence of a background signal which poses a physical limit. Indeed, this background comes with its own counting noise which can swamp the signal of interest from e.g. a low concentration element in a sample in many cases of interest\cite{Reimer2008, Shuman1986}. For these reasons it is commonly accepted that EELS is preferred for low Z elements at not too low concentrations where it provides a rich amount of fine spectroscopic data and EDX is preferred for heavier elements and good detection efficiency (very little or no background and peaks are often well separated in energy).\par
\begin{figure}[t!]
    \includegraphics[width=0.5\textwidth]{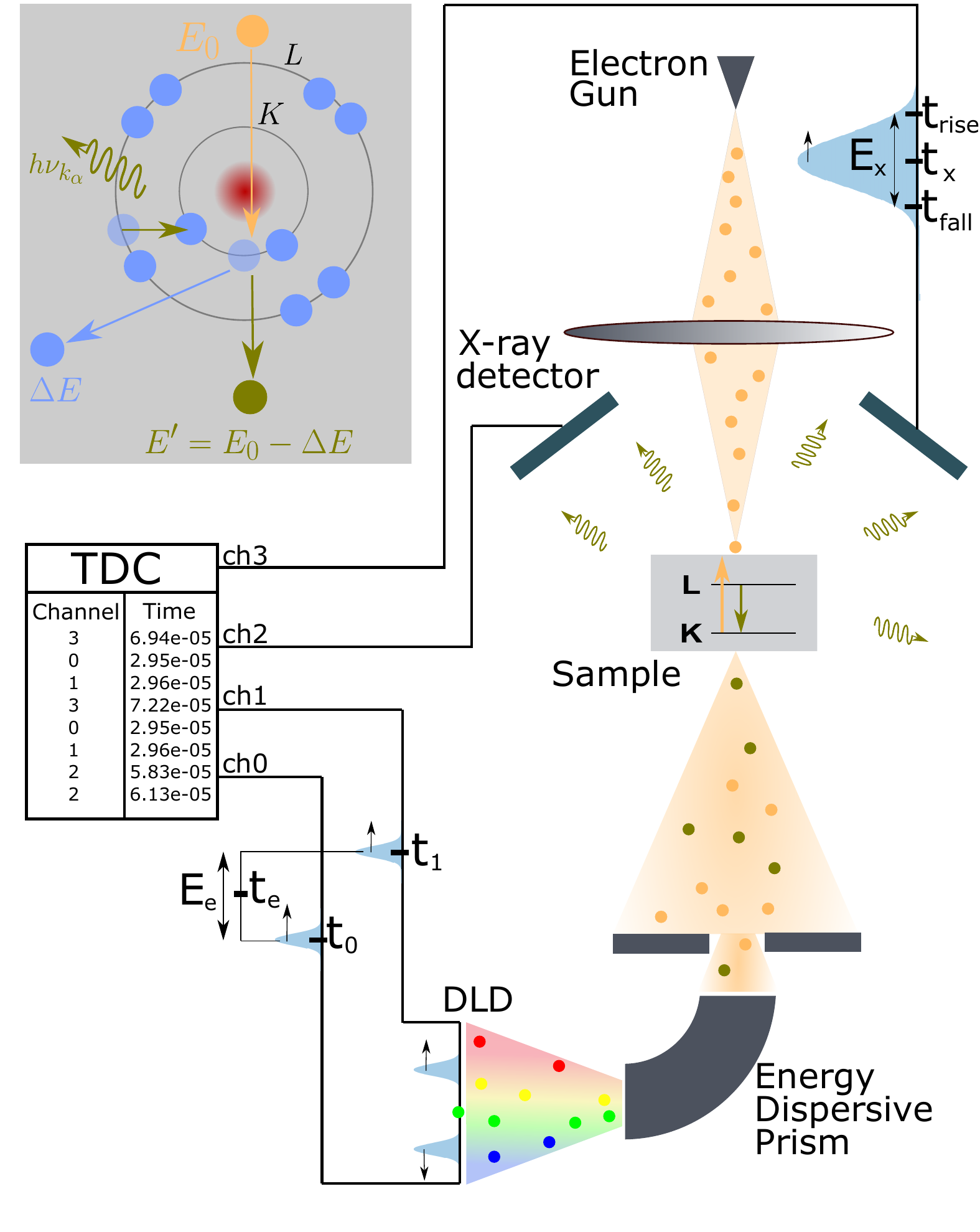}
	\caption{Sketch of the implemented coincidence detection setup. The incoming electron creates a inner-shell excitation which decays emitting an X-ray. At the top left a visualization of this process is shown. The correlated X-rays and electrons are indicated with the same color. Both events (electrons and X-rays) are detected and their energy and time of occurrence are determined after processing the data stream of the TDC (time-to-digital converter).}
\label{microscope_setup}
\end{figure}
\begin{figure}[t]
	\includegraphics[width=0.5\textwidth]{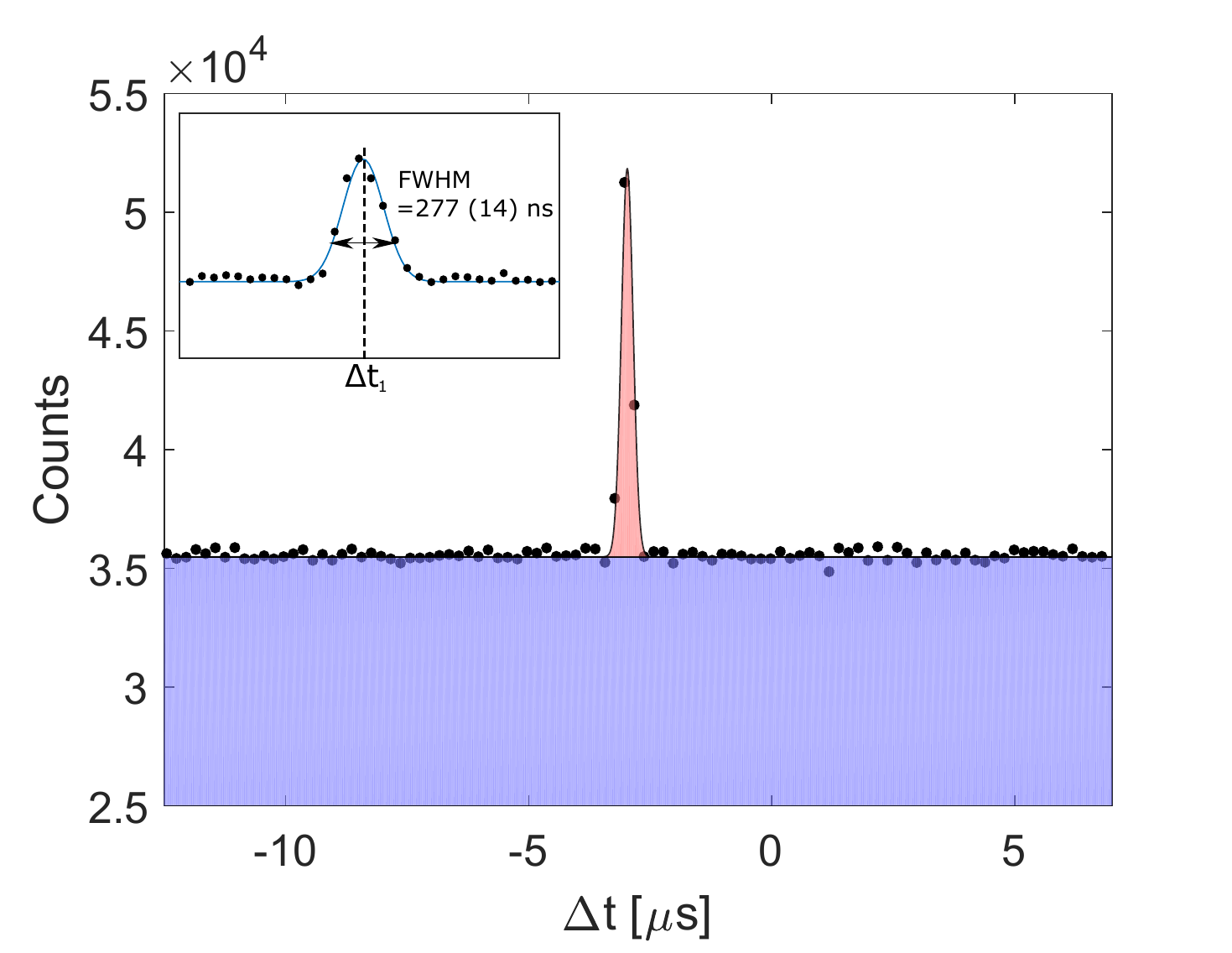}
	\caption{Histogram of the time difference $\Delta$t between X-ray and EEL events showing a clear temporal correlation (red peak). Uncorrelated events are present as a background (blue). The width of the coincidence peak (inset), is assumed to be predominantly instrumental in nature.}
	\label{proof}
\end{figure}
\begin{figure*}[t]
	\includegraphics[width=0.8\textwidth]{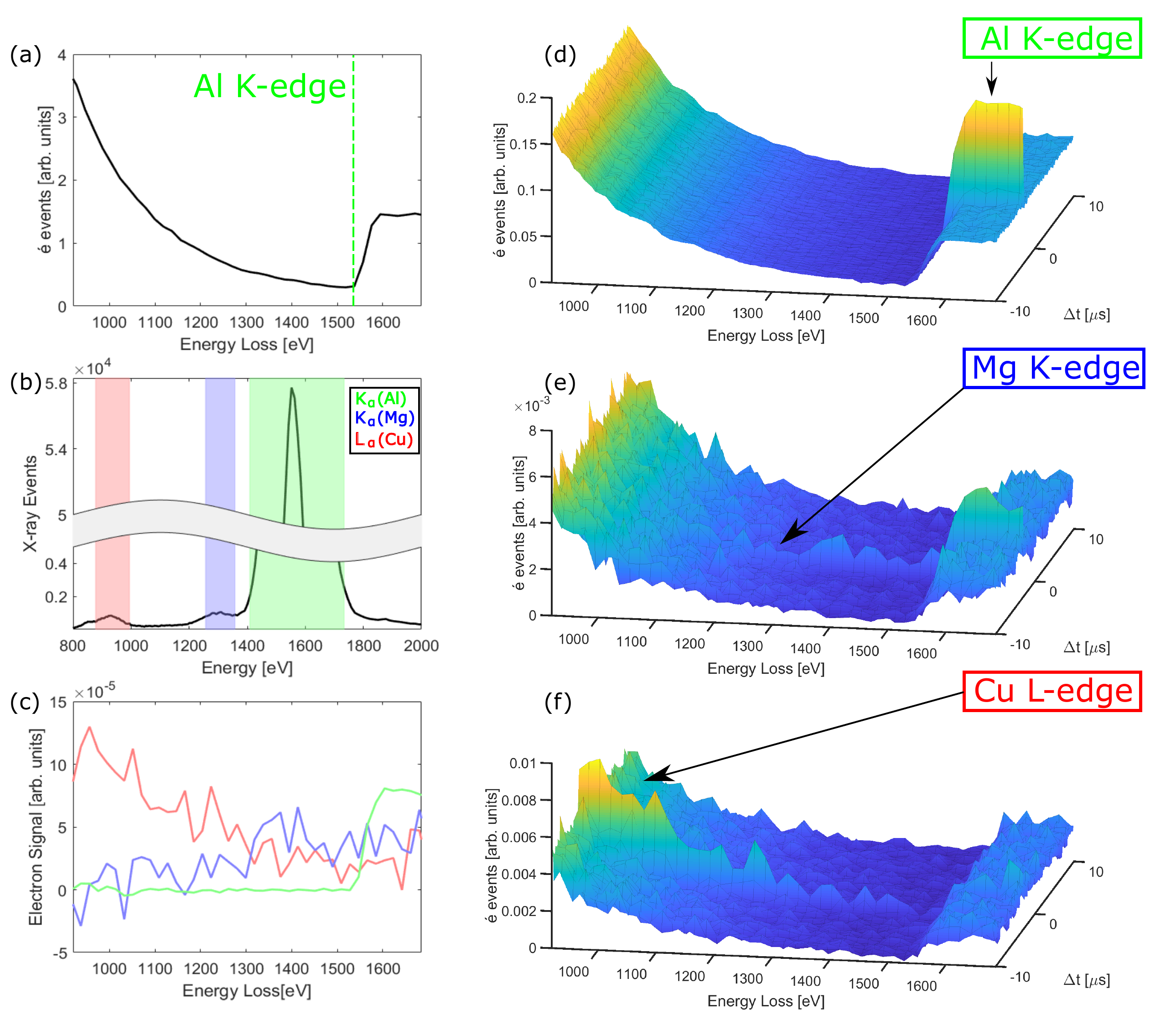}
	\caption{
		(a) The EEL spectrum of the Al-Mg-Si-Cu alloy where the aluminium K-edge is indicated. (b) The X-ray spectrum where three X-rays originating from different atoms are marked. (d-f) The post selection of electron events, when different energy windows of X-ray energies are selected, as a function of the electron energy loss and time difference between the X-ray and electron event. The selected energy windows correspond to the characteristic X-ray energies of the different atoms (Al, Mg, Cu) present in the sample. It is clear that at a particular time difference there is a increase in signal. This increased signal corresponds to the core-level ionization event followed by X-ray emission with an energy inside the selected window. (c) The RUP subtracted coincidence EEL spectra for the different elements where the colour-code is identical as for figure (b).
	}	
	\label{2deels}
\end{figure*}
The experiment was performed on an FEI Tecnai Osiris operated at 200~keV equipped with a Super-X EDX detector and a Gatan Imaging Filter (GIF 200). The EDX detection setup consists of four Silicon Drift Detectors (SDD) providing a collection solid angle of approximately 0.7~sr. However, only two of four detectors were used due to limitations in our current hardware setup but future improvements should allow to use all four detectors. A microchannel plate (MCP) coupled to a  delay line detector\cite{Oelsner2001,Muller-Caspary2015a} (DLD) was mounted at the back of the GIF200 EEL spectrometer, replacing the standard CCD camera. The DLD has a timing resolution of $\sim$ 30~ps which is more than sufficient to reveal the coincidence information. An \ce{Al-Mg-Si-Cu} alloy\cite{Lipeng2018} was selected as a test sample due to rich features in both EELS and EDX spectra. Furthermore, the alloy is challenging for conventional EELS quantification because of the relatively low abundance of \ce{Mg} ($\approx$ 0.62 wt\%), \ce{Si} ($\approx$ 1.11 wt\%) and \ce{Cu} ($\approx$ 0.5 wt\%). Additionally, there is the difficulty of detecting \ce{Cu} from EDX due to multiple sample unrelated sources of \ce{Cu} in several components of the microscope and the supporting grid. We will show that both difficulties can be overcome making use of the coincidence information, allowing for improved quantification.\par
In Fig.\ref{microscope_setup} a sketch of the coincidence detection setup is shown, depicting the excitation via inelastic electron scattering to an atom in the sample which undergoes a core-level ionization. Next, the excited atom can decay by emitting an X-ray while filling up the vacancy created in the EEL event. The electron and X-ray are detected by the DLD and SDD, respectively. In order to measure the arrival time and energy of the EDX signal, we use a comparator circuit to the two (out of four) analog outputs of the pulse shaping hardware of the SDD in order to record the rising and falling edge of this pulse through external inputs on the time-to-digital converter (TDC). The energy of the X-ray event then relates to the time difference between the rising and falling edge while the arrival time is defined as the mean of the rising and falling edge (see Fig.\ref{microscope_setup}). The DLD detector records the coordinates of every detected electron by measuring the time difference of a pulse generated by the electron interacting with the MCP traveling towards both ends of a finite length wire (the delay line). These coordinates relate to the energy of the incoming electron due to the energy dispersive nature of the EEL spectrometer (see Fig.\ref{microscope_setup}). The arrival time of the electron is taken as the mean time of both pulses $t_e = (t_{2}+t_{1})/2$.\par
Using this experimental setup, the energy and arrival time is obtained and stored for every detected electron ($E_e$,$t_e$) and X-ray ($E_x$,$t_x$). In order to obtain information on the time correlation, the time differences between EELS and EDX events are important, not the absolute arrival time of the events. Therefore for every EEL event, the time difference is calculated between this EEL event and every EDX event. However, only time differences in the interval of [-20 ,20]~$\mu$s are selected since any correlation is expected to occur almost instantaneous and well inside this timing interval. For every time difference also the corresponding energies of the EEL and EDX are stored as a list consisting of EEL energy, EDX energy and time difference between these two events ($E_e$,$E_x$,$\Delta t$).\par 
The events classified as ($E_e$,$E_x$,$\Delta t$) provide all the information needed for observing the time coincidences between the electrons and X-rays. The first step is to verify that at $\Delta$t = 0 there is an increase in counts as would be expected if both events are correlated. The histogram of the time difference between EELS and EDX events is given in Fig.\ref{proof} and shows a clear coincidence peak at $\Delta$t$_1 \approx 3~\mu$s, corresponding to the inelastic scattering of an electron exciting an atom followed by the decay of the same atom via an X-ray emission process. The delay of 3~$\mu$s was unexpected since the duration of the inner-shell excitation and de-excitation by X-ray emission is expected to be in the order of femtoseconds\cite{Krause1979}. This time delay is likely due to the analog pulse shaping process of the EDX signal and can further be ignored as an offset that does not influence the analysis. The width of the peak in Fig.\ref{proof} is related to the time selectivity $\tau$ of the coincidence setup which is estimated here to be 277(14)~ns. This is significantly higher than the expected drift time in the SDD detectors and could be related to noise in the comparator circuit leading to unwanted timing jitter which should be further improved on in future versions of the setup.\par 
In addition to the coincidence peak there is a relatively large unspecific background which originates from random uncorrelated processes (RUP) happening inside a given time interval. For instance, an electron excites an atom but in the same time interval another atom emits an X-ray, then this first electron and X-ray are both detected and form the constant background indicated in blue. We can estimate this RUP signal quite accurately by measuring all events outside the coincidence window $\Delta t_1 \pm \tau/2$.\par
Fig.\ref{2deels}(a) and (b) shows the EEL and EDX spectrum of the alloy sample respectively. In the EDX spectrum, the characteristic X-ray energy windows from different atoms (K$_\alpha$(\ce{Al}), K$_\alpha$(\ce{Mg}), L$_\alpha$(\ce{Cu})) are marked. In Fig.\ref{2deels}(e-f) EEL events are represented as a function of energy loss and time difference when different energy windows of X-ray energies are selected. We observe that at $\Delta$t$_1$ the element specific signal increases significantly, depending on the selected X-ray energy window. 
This already shows the potential and selectivity of the technique as also the very weak \ce{Cu} L-edge and \ce{Mg} K-edge are appearing while they are totally indistinguishable from the background in Fig.\ref{2deels}(a). Furthermore, the signal outside $\Delta$t$_1$ is constant as a function of time difference which is expected because these RUP events  have no correlation in time. This means that the signal inside the time coincidence window can be thought of as the sum between the RUP events and the true correlated events. Additionally, since the electron energy distribution of the RUP events is known accurately due to the high statistics, this signal can be scaled and subtracted from the signal inside the time coincidence window.\par
As a result, only the true core-level ionization events remain as demonstrated in Fig.\ref{2deels}(c) while the background signal is entirely removed. This background removal is very attractive as compared to conventional extrapolation of a power law function from a fitting region before the edge. Here we need no fitting region, no assumptions on the shape of the background and we do not suffer from extrapolation errors that can easily become larger than the edge signal for low-concentration elements. 
In order to demonstrate this, we compare conventional quantification making use of such an extrapolated background with results obtained when using coincidence as described above. The results are summarised in table \ref{quantificat} (see supp. material for details). The values for the coincidence setup have a markedly better precision and agree with the values obtained by Lipeng et al. using different methods \cite{Lipeng2018}. This demonstrates that coincidence detection of EELS and EDX provides a substantial advantage over conventional EELS as it allows to selectively boost the signal to background for specific excitation edges, revealing a much better detectability, especially for weak edges, while avoiding the conventional background fitting and removal step. This means that here we can even quantify the \ce{Cu} L-edge  although the spectrum contains no energies prior to the edge and therefore conventional quantification is entirely impossible.
Compared to EDX quantification, the coincidence technique has the benefit that no unwanted fluorescence signal is present as such X-ray events would not be accompanied by the required EEL event to obtain coincidence. This solves a major issue in conventional EDX quantification and is especially important for \ce{Cu} here.\par
\begin{table}
	\begin{center}
		\begin{tabular}{c | c | c | c }
			Element &Coincidence &EDX&EELS\\
			 
			\hline
			\ce{Cu}&2.65$\pm$ 0.07&3.0$\pm$0.3& not possible\\
			\ce{Mg}&0.61$\pm$ 0.04&1.42$\pm$0.08&7.45$\pm$0.5
		\end{tabular}
		\caption{Elemental quantification [wt\%] of magnesium and copper in the aluminum alloy matrix obtained with coincidence detection and compared to conventional quantification of EDX and EELS (for details see supp. information).}
		\label{quantificat}	
	\end{center}
\end{table}
Note that in principle, the coincidence timing peak could be further narrowed with improved electronics. This would result in an almost completely suppressed background without having to subtract the RUP, as the influence of the RUP decreases when the time resolution improves. It is important to note that, even though many EEL events are discarded in the coincidence filtering, the signal to noise ratio of the signal of interest does actually improve as the signal to background ratio is significantly increased. This is the essential reason for the superior precision estimates as compared to conventional quantification methods for either EDX or EELS while at the same time, significant sources of systematic error in both EELS (background extrapolation) or EDX (fluorescence) are overcome improving the accuracy of quantification.\par
We demonstrate how new detector developments lead to rich datasets that provide more than the simple correlation of two signals while still collecting all detector events. In particular, a time coincidence setup is demonstrated using a DLD and SDD detector storing information about the energy and arrival time of every inelastic electron and X-ray event. 
This method provides model free background removal for EELS and solves fluorescence issues in EDX. Both precision and accuracy of quantification can be significantly improved, perhaps surprisingly, by filtering out unwanted events. By storing all events and their time of arrival, a wide range of post processing and pattern recognition options becomes available. Storing event data this way, has the intrinsic benefit of natural data compression, where only actual events are stored, which especially for low countrate signals can be an improvement over more traditional recording schemes. 
The proposed setup can further be improved by increasing time resolution of the EDX setup using different electronics and incorporating four SDD detectors. Although, improvements are possible, the results demonstrate the usefulness of time stamped event detection in the TEM and indicate that technology is ready to provide this rich source of information.\par
In this paper we focussed on timing coincidences between EELS and EDX events but the idea is general and can be extended to e.g. EELS and Auger electrons\cite{VOREADES1976,Mullejans1993}, EELS and cathode-luminescence (CL) \cite{Graham1986} and even for quantum coincidence measurements between two CL events showing photon-anti-bunching in diamond Nitrogen-vacancy (NV) centers\cite{Meuret2015}.
Such coincidence experiments are typical for particle accelerators where timing and trajectory information is often superior to what is possible in TEM, but spatial resolution and beam coherence are typically rather poor compared to modern TEM instruments. Making use of the advanced detectors developed for particle accelerator experiments, we can now combine the benefits of both instruments to provide new capabilities\cite{Oelsner2001, Kremer1988}.\par
\newpage
D.J., A.B. and J.V. acknowledge funding from FWO project G093417N. This project has received funding from the European Union's Horizon 2020 research and innovation programme under grant agreement No 823717 – ESTEEM3. The authors acknowledge financial support from the Research Fund of the University of Antwerp. Provision of parts of the DLD electronics by Prof. Andreas Rosenauer (Bremen University, Germany) is gratefully acknowledged. KMC acknowledges funding from grant VH-NG-1317 of the Helmholtz association.\par
The data and data processing can be found at doi:10.5281/zenodo.2563880. \par

\section{Supplementary Information: Elemental quantification of the Al-Mg-Si-Cu alloy}
In this supplementary information we address details of the quantification procedure of the Al-Mg-Si-Cu alloy
\subsection{Quantification conventional EELS}
In the conventional EELS spectrum (Fig.\ref{background}(a)) the aluminium K edge is clearly visible. The \ce{Mg} edge, on the other hand, is not immediately discernible due to the low concentration of this element and the high background signal. Additionally, the \ce{Cu} is not visible because the spectrum contains no energies prior to the edge which makes conventional quantification impossible. On the other hand, a guess for the quantification of \ce{Al} and \ce{Mg} can be done. The analysis was performed using the open-source software
package HyperSpy \cite{Pena2018}. For the \ce{Mg}, the background was fitted using a powerlaw fit and the pre-edge was chosen from 1000 eV to 1300 eV. The interval used for the quantification was from 1300 eV to 1350 eV. For the \ce{Al} a pre-edge window of 1300-1525 eV was chosen and the interval used for quantification was from 1600-1700 eV. The wt\% was calculated as follows:
\begin{equation}
\ce{wt}\%(\ce{Mg})=\frac{N(\ce{Mg}) \sigma(\ce{Al})}{N(\ce{Al}) \sigma(\ce{Mg})}  
\label{eels} 
\end{equation}
The result is shown in Table \ref{quantificat}. The obtained value has a large deviation compared to the EDX and coincidence. 	
\subsection{Quantification conventional EDX}
The quantification of the conventional EDX spectrum was performed using the ESPRIT software version 1.9. The standard method using the Cliff-Lorimer equation is used. The results are shown in Table \ref{quantificat}. 
\subsection{Quantification coincidence EELS-EDX}
\begin{figure*}[t]
	\includegraphics[width=0.95\textwidth]{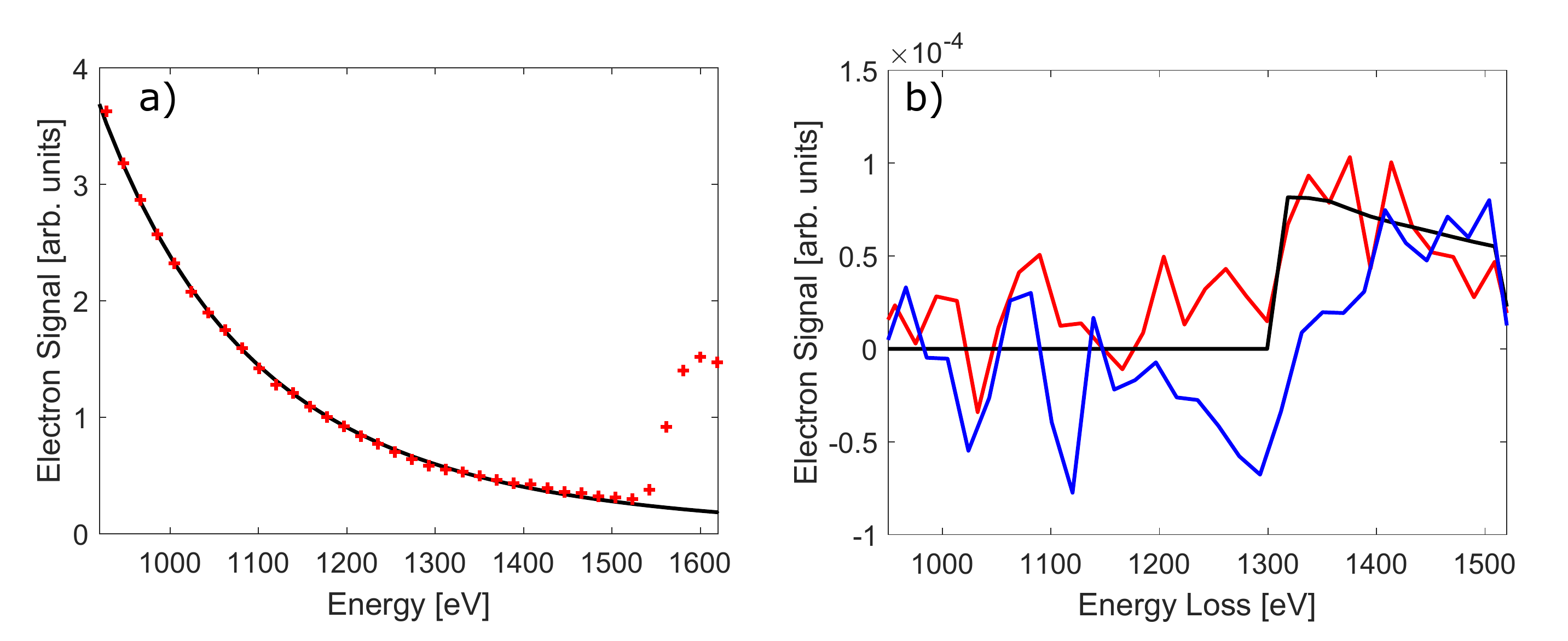}
	\caption{(a) The conventional EELS spectrum (red crosses) where the background (black) before the magnesium edge is fitted with a power law. (b) The conventional background subtracted EELS spectrum (blue), the background subtracted coincidence EELS spectrum (red) and the cross section for the K-edge of magnesium. The conventional spectrum looks like noise only and inside K-edge energy interval the trend is not similar to the cross section. For the coincidence EELS there is also substantial noise. However inside the K-edge energy interval, the spectrum follows the trend of the cross section.}
	\label{background}
\end{figure*}

\begin{figure}[h]
	\includegraphics[width=0.5\textwidth]{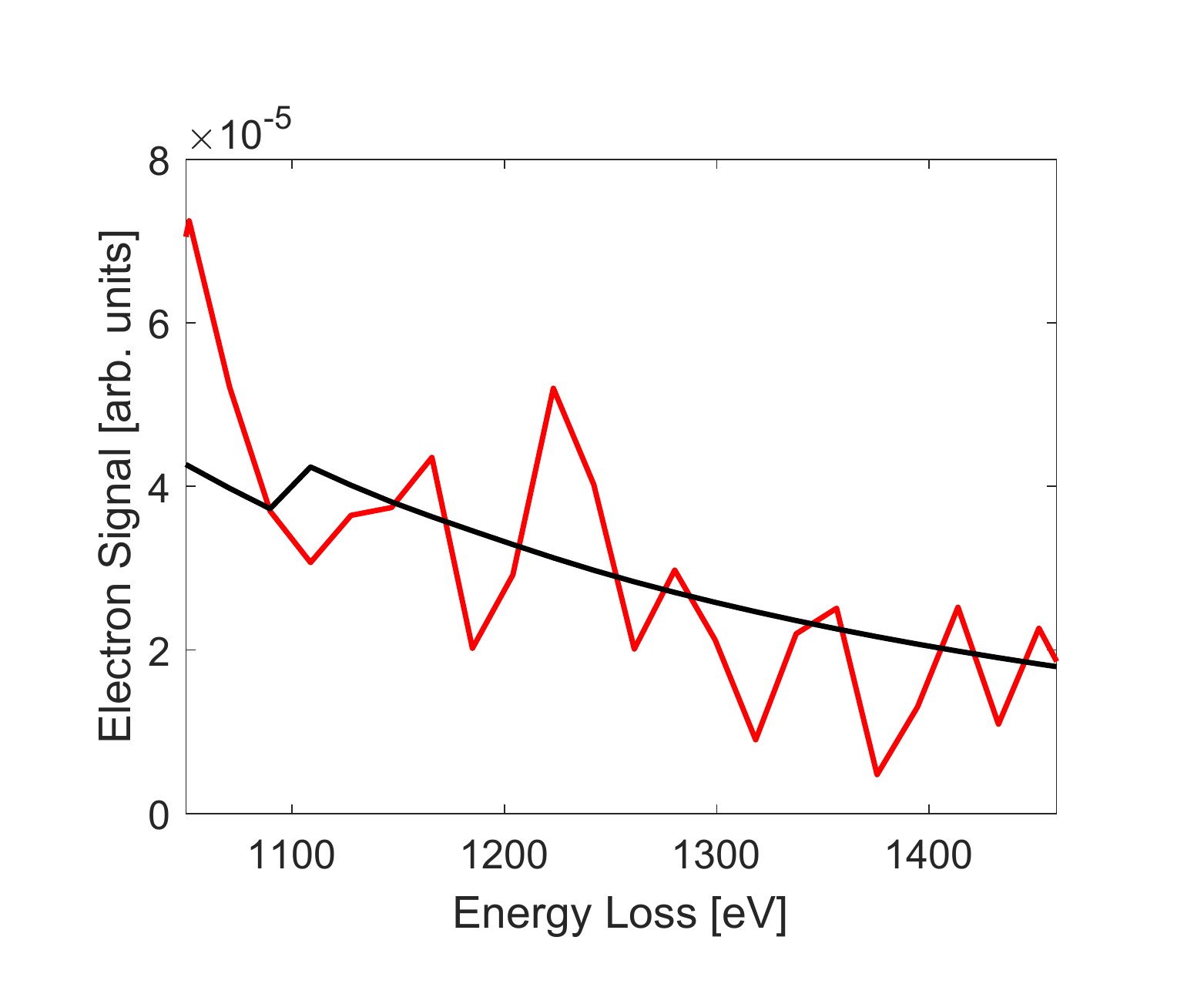}
	\centering
	\caption{The copper L-edge in the coincidence method (red) and the theoretical cross section (black). The experimental data follows the same trend as the theoretical cross section.}
	\label{copperedge}
\end{figure}

When applying coincidence post selection however, we notice that the signal represented in Fig.\ref{background}(c), becomes identifiable, perhaps surprisingly by deliberately throwing away EELS events that were not accompanied by an X ray of the right energy range. These edges can now be used to quantify the wt\% of the \ce{Cu} and \ce{Mg} in the \ce{Al} matrix. The first steps of this process is the post-selection of the X-ray energy windows for the different elements.
Additionally, the time window in which the coincidence has to occur is set to -3.15$\leq$$\Delta$t$\leq$-2.85 $\mu$s . The next step is the subtraction of the background signal (all energy selected events that were outside the time window) from the coincidence signal (inside the time window). The result is a spectrum containing only post selected events with no background which can be readily quantified (Fig.\ref{copperedge}). The number of counts (N) for element A in the coincidence EELS spectrum as a function of energy loss is given by following equation:
\begin{equation}
N_{A}(E)=C_{A} \cdot \epsilon_{x}\epsilon_e \cdot \omega_{A} \cdot  \sigma_{A}(E) \cdot \frac{I}{M_{A}}  
\label{count} 
\end{equation}
where M is the atomic mass, $\epsilon_{x}$, $\epsilon_e$ are the efficiencies of the EDX and EELS setup respectively, $\omega$ is the fluorescence yield, $\sigma(E)$ is the differential cross section at energy E, I is the incoming electron current and C is the mass concentration. Note that this formula holds when in the X-ray window the entire peak for the particular element is selected. Moreover, Eq. (\ref{count}) does not take X-ray absorption into account and it also assumes that the efficiency of the EDX setup is independent on X-ray energy which is almost true if the characteristic X-rays have similar energies. Although there are a number of assumptions it is still valid to approximate the abundances using Eq. (\ref{count}). The cross section of the different edges is determined using the experimental parameters (electron energy $200$~keV, convergence semi angle 20~mrad and collection semi angle 100~mrad). The fluorescence yield of the various elements is found in the literature \cite{Hubbel2004}. In order to determine the wt\%, for each element, an energy window in EELS is chosen for each element. The wt\% of Mg and Cu are calculated by taking the ratio of the concentrations. In this way it is clear that the weight percentage is independent on the efficiencies and the electron current.
\begin{equation}
\ce{wt}\%(A)=\frac{C_{A}}{C_{B}} = \frac{M_{A} \cdot \omega_{B}}{M_{B} \cdot \omega_{B}} \frac{\sum_{i} N_{A}(E_i)}{\sum_{i}\sigma_{A}(E_i)} \cdot \frac{\sum_{j}\sigma_{B}(E_j)}{\sum_{j} N_{B}(E_j)}
\label{quant}
\end{equation}
\begin{equation}
\ce{wt}\%(A)= k_{AB} \cdot \frac{\sum_{i}N_{A}(E_i)}{\sum_{j}N_{B}(E_j)}
\end{equation}
By rewriting the previous equation, it is seen that the weight percentage depends on a k-factor multiplied with the ratio of the number of counts in the coincidence spectrum. This is similar to the Cliff-Lorimer equation which is well known in conventional EDX quantification.
It should be noted that in formula \ref{quant} some approximations were applied. However, in the future these could be surpassed by experimentally determining the k-factors as is done in conventional EELS and EDX. Using the above procedure we find the following quantification results.
The values calculated are comparable to the ones given in Lipeng et al. \cite{Lipeng2018}. The small deviations could arise due to the high spatial resolution of the setup where only a small part of the material is probed and hence the relative abundances can deviate from the average values. Table \ref{quantificat} shows that relatively low abundances can be detected (in principle) at high spatial and energy resolution which was not possible in conventional EELS when no post selection was applied. 
Additionally, in the coincidence method there was no need for background fitting which adds complications to the quantification process. From the point of view of EDX there is an additional advantage which is due to the fact that in the coincidence technique the signal coming from fluorescence X-rays are in the background since they are not considered as proper coincidences (X-ray is not in the right energy window). Thereby this technique improves the precision of quantification.

\newpage
\newpage

\bibliography{biblio_1}

\begin{thebibliography}{25}%
\makeatletter
\providecommand \@ifxundefined [1]{%
 \@ifx{#1\undefined}
}%
\providecommand \@ifnum [1]{%
 \ifnum #1\expandafter \@firstoftwo
 \else \expandafter \@secondoftwo
 \fi
}%
\providecommand \@ifx [1]{%
 \ifx #1\expandafter \@firstoftwo
 \else \expandafter \@secondoftwo
 \fi
}%
\providecommand \natexlab [1]{#1}%
\providecommand \enquote  [1]{``#1''}%
\providecommand \bibnamefont  [1]{#1}%
\providecommand \bibfnamefont [1]{#1}%
\providecommand \citenamefont [1]{#1}%
\providecommand \href@noop [0]{\@secondoftwo}%
\providecommand \href [0]{\begingroup \@sanitize@url \@href}%
\providecommand \@href[1]{\@@startlink{#1}\@@href}%
\providecommand \@@href[1]{\endgroup#1\@@endlink}%
\providecommand \@sanitize@url [0]{\catcode `\\12\catcode `\$12\catcode
  `\&12\catcode `\#12\catcode `\^12\catcode `\_12\catcode `\%12\relax}%
\providecommand \@@startlink[1]{}%
\providecommand \@@endlink[0]{}%
\providecommand \url  [0]{\begingroup\@sanitize@url \@url }%
\providecommand \@url [1]{\endgroup\@href {#1}{\urlprefix }}%
\providecommand \urlprefix  [0]{URL }%
\providecommand \Eprint [0]{\href }%
\providecommand \doibase [0]{http://dx.doi.org/}%
\providecommand \selectlanguage [0]{\@gobble}%
\providecommand \bibinfo  [0]{\@secondoftwo}%
\providecommand \bibfield  [0]{\@secondoftwo}%
\providecommand \translation [1]{[#1]}%
\providecommand \BibitemOpen [0]{}%
\providecommand \bibitemStop [0]{}%
\providecommand \bibitemNoStop [0]{.\EOS\space}%
\providecommand \EOS [0]{\spacefactor3000\relax}%
\providecommand \BibitemShut  [1]{\csname bibitem#1\endcsname}%
\let\auto@bib@innerbib\@empty
\bibitem [{\citenamefont {Cappellen}\ and\ \citenamefont
  {Doukhan}(1994)}]{VanCappellen1994}%
  \BibitemOpen
  \bibfield  {author} {\bibinfo {author} {\bibfnamefont {E.~V.}\ \bibnamefont
  {Cappellen}}\ and\ \bibinfo {author} {\bibfnamefont {J.~C.}\ \bibnamefont
  {Doukhan}},\ }\bibfield  {title} {\enquote {\bibinfo {title} {Quantitative
  transmission x-ray microanalysis of ionic compounds},}\ }\href {\doibase
  https://doi.org/10.1016/0304-3991(94)90047-7} {\bibfield  {journal} {\bibinfo
   {journal} {Ultramicroscopy}\ }\textbf {\bibinfo {volume} {53}},\ \bibinfo
  {pages} {343 -- 349} (\bibinfo {year} {1994})}\BibitemShut {NoStop}%
\bibitem [{\citenamefont {Williams}\ and\ \citenamefont
  {Carter}(2009)}]{Williams2009}%
  \BibitemOpen
  \bibfield  {author} {\bibinfo {author} {\bibfnamefont {D.~B.}\ \bibnamefont
  {Williams}}\ and\ \bibinfo {author} {\bibfnamefont {C.~B.}\ \bibnamefont
  {Carter}},\ }\href@noop {} {\emph {\bibinfo {title} {Transmission Electron
  Microscopy - A Textbook for Materials Science}}},\ edited by\ \bibinfo
  {editor} {\bibfnamefont {D.~B.}\ \bibnamefont {Williams}}\ and\ \bibinfo
  {editor} {\bibfnamefont {C.~B.}\ \bibnamefont {Carter}}\ (\bibinfo
  {publisher} {Springer},\ \bibinfo {year} {2009})\BibitemShut {NoStop}%
\bibitem [{\citenamefont {Schlossmacher}\ \emph {et~al.}(2010)\citenamefont
  {Schlossmacher}, \citenamefont {Klenov}, \citenamefont {Freitag},\ and\
  \citenamefont {von Harrach}}]{Schlossmacher2010}%
  \BibitemOpen
  \bibfield  {author} {\bibinfo {author} {\bibfnamefont {P.}~\bibnamefont
  {Schlossmacher}}, \bibinfo {author} {\bibfnamefont {D.}~\bibnamefont
  {Klenov}}, \bibinfo {author} {\bibfnamefont {B.}~\bibnamefont {Freitag}}, \
  and\ \bibinfo {author} {\bibfnamefont {H.}~\bibnamefont {von Harrach}},\
  }\bibfield  {title} {\enquote {\bibinfo {title} {Enhanced detection
  sensitivity with a new windowless xeds system for aem based on silicon drift
  detector technology},}\ }\href {\doibase 10.1017/S1551929510000404}
  {\bibfield  {journal} {\bibinfo  {journal} {Microscopy Today}\ }\textbf
  {\bibinfo {volume} {18}},\ \bibinfo {pages} {14--20} (\bibinfo {year}
  {2010})}\BibitemShut {NoStop}%
\bibitem [{\citenamefont {Egerton}(2011)}]{Egerton2011}%
  \BibitemOpen
  \bibfield  {author} {\bibinfo {author} {\bibfnamefont {R.}~\bibnamefont
  {Egerton}},\ }\href@noop {} {\emph {\bibinfo {title} {Electron Energy-Loss
  Spectroscopy in the Electron Microscope}}}\ (\bibinfo {year}
  {2011})\BibitemShut {NoStop}%
\bibitem [{\citenamefont {Tan}\ \emph {et~al.}(2011)\citenamefont {Tan},
  \citenamefont {Turner}, \citenamefont {Y\"ucelen}, \citenamefont {Verbeeck},\
  and\ \citenamefont {Van~Tendeloo}}]{Tan2011}%
  \BibitemOpen
  \bibfield  {author} {\bibinfo {author} {\bibfnamefont {H.}~\bibnamefont
  {Tan}}, \bibinfo {author} {\bibfnamefont {S.}~\bibnamefont {Turner}},
  \bibinfo {author} {\bibfnamefont {E.}~\bibnamefont {Y\"ucelen}}, \bibinfo
  {author} {\bibfnamefont {J.}~\bibnamefont {Verbeeck}}, \ and\ \bibinfo
  {author} {\bibfnamefont {G.}~\bibnamefont {Van~Tendeloo}},\ }\bibfield
  {title} {\enquote {\bibinfo {title} {2d atomic mapping of oxidation states in
  transition metal oxides by scanning transmission electron microscopy and
  electron energy-loss spectroscopy},}\ }\href {\doibase
  10.1103/PhysRevLett.107.107602} {\bibfield  {journal} {\bibinfo  {journal}
  {Phys. Rev. Lett.}\ }\textbf {\bibinfo {volume} {107}},\ \bibinfo {pages}
  {107602} (\bibinfo {year} {2011})}\BibitemShut {NoStop}%
\bibitem [{\citenamefont {Tan}\ \emph {et~al.}(2012)\citenamefont {Tan},
  \citenamefont {Verbeeck}, \citenamefont {Abakumov},\ and\ \citenamefont
  {Tendeloo}}]{Tan2012}%
  \BibitemOpen
  \bibfield  {author} {\bibinfo {author} {\bibfnamefont {H.}~\bibnamefont
  {Tan}}, \bibinfo {author} {\bibfnamefont {J.}~\bibnamefont {Verbeeck}},
  \bibinfo {author} {\bibfnamefont {A.}~\bibnamefont {Abakumov}}, \ and\
  \bibinfo {author} {\bibfnamefont {G.~V.}\ \bibnamefont {Tendeloo}},\
  }\bibfield  {title} {\enquote {\bibinfo {title} {Oxidation state and chemical
  shift investigation in transition metal oxides by eels},}\ }\href {\doibase
  https://doi.org/10.1016/j.ultramic.2012.03.002} {\bibfield  {journal}
  {\bibinfo  {journal} {Ultramicroscopy}\ }\textbf {\bibinfo {volume} {116}},\
  \bibinfo {pages} {24 -- 33} (\bibinfo {year} {2012})}\BibitemShut {NoStop}%
\bibitem [{\citenamefont {Turner}\ \emph {et~al.}(2012)\citenamefont {Turner},
  \citenamefont {Verbeeck}, \citenamefont {Ramezanipour}, \citenamefont
  {Greedan}, \citenamefont {Van~Tendeloo},\ and\ \citenamefont
  {Botton}}]{Turner2012}%
  \BibitemOpen
  \bibfield  {author} {\bibinfo {author} {\bibfnamefont {S.}~\bibnamefont
  {Turner}}, \bibinfo {author} {\bibfnamefont {J.}~\bibnamefont {Verbeeck}},
  \bibinfo {author} {\bibfnamefont {F.}~\bibnamefont {Ramezanipour}}, \bibinfo
  {author} {\bibfnamefont {J.~E.}\ \bibnamefont {Greedan}}, \bibinfo {author}
  {\bibfnamefont {G.}~\bibnamefont {Van~Tendeloo}}, \ and\ \bibinfo {author}
  {\bibfnamefont {G.~A.}\ \bibnamefont {Botton}},\ }\bibfield  {title}
  {\enquote {\bibinfo {title} {Atomic resolution coordination mapping in
  ca2fecoo5 brownmillerite by spatially resolved electron energy-loss
  spectroscopy},}\ }\href {\doibase 10.1021/cm300640g} {\bibfield  {journal}
  {\bibinfo  {journal} {Chemistry of Materials}\ }\textbf {\bibinfo {volume}
  {24}},\ \bibinfo {pages} {1904--1909} (\bibinfo {year} {2012})},\ \Eprint
  {http://arxiv.org/abs/https://doi.org/10.1021/cm300640g}
  {https://doi.org/10.1021/cm300640g} \BibitemShut {NoStop}%
\bibitem [{\citenamefont {Muller D.~A.}(1999)}]{Muller1999}%
  \BibitemOpen
  \bibfield  {author} {\bibinfo {author} {\bibfnamefont {M.~S. B. F. H. E.-L.
  K. T.~G.}\ \bibnamefont {Muller D.~A.}, \bibfnamefont {Sorsch~T.}},\
  }\bibfield  {title} {\enquote {\bibinfo {title} {{The electronic structure at
  the atomic scale of ultrathin gate oxides}},}\ }\href {\doibase
  10.1038/21602} {\bibfield  {journal} {\bibinfo  {journal} {Nature}\ }\textbf
  {\bibinfo {volume} {399}},\ \bibinfo {pages} {3967--3969} (\bibinfo {year}
  {1999})}\BibitemShut {NoStop}%
\bibitem [{\citenamefont {Dudarev}\ \emph {et~al.}(1998)\citenamefont
  {Dudarev}, \citenamefont {Botton}, \citenamefont {Savrasov}, \citenamefont
  {Humphreys},\ and\ \citenamefont {Sutton}}]{Dudarev1998}%
  \BibitemOpen
  \bibfield  {author} {\bibinfo {author} {\bibfnamefont {S.~L.}\ \bibnamefont
  {Dudarev}}, \bibinfo {author} {\bibfnamefont {G.~A.}\ \bibnamefont {Botton}},
  \bibinfo {author} {\bibfnamefont {S.~Y.}\ \bibnamefont {Savrasov}}, \bibinfo
  {author} {\bibfnamefont {C.~J.}\ \bibnamefont {Humphreys}}, \ and\ \bibinfo
  {author} {\bibfnamefont {A.~P.}\ \bibnamefont {Sutton}},\ }\bibfield  {title}
  {\enquote {\bibinfo {title} {Electron-energy-loss spectra and the structural
  stability of nickel oxide: An lsda+u study},}\ }\href {\doibase
  10.1103/PhysRevB.57.1505} {\bibfield  {journal} {\bibinfo  {journal} {Phys.
  Rev. B}\ }\textbf {\bibinfo {volume} {57}},\ \bibinfo {pages} {1505--1509}
  (\bibinfo {year} {1998})}\BibitemShut {NoStop}%
\bibitem [{\citenamefont {Ali}\ \emph {et~al.}(1999)\citenamefont {Ali},
  \citenamefont {D{\"o}rner}, \citenamefont {Jagutzki}, \citenamefont
  {N{\"u}ttgens}, \citenamefont {Mergel}, \citenamefont {Spielberger},
  \citenamefont {Khayyat}, \citenamefont {Vogt}, \citenamefont {Br{\"a}uning},
  \citenamefont {Ullmann}, \citenamefont {Moshammer}, \citenamefont {Ullrich},
  \citenamefont {Hagmann}, \citenamefont {Groeneveld}, \citenamefont {Cocke},\
  and\ \citenamefont {Schmidt-B{\"o}cking}}]{Ali1999}%
  \BibitemOpen
  \bibfield  {author} {\bibinfo {author} {\bibfnamefont {I.}~\bibnamefont
  {Ali}}, \bibinfo {author} {\bibfnamefont {R.}~\bibnamefont {D{\"o}rner}},
  \bibinfo {author} {\bibfnamefont {O.}~\bibnamefont {Jagutzki}}, \bibinfo
  {author} {\bibfnamefont {S.}~\bibnamefont {N{\"u}ttgens}}, \bibinfo {author}
  {\bibfnamefont {V.}~\bibnamefont {Mergel}}, \bibinfo {author} {\bibfnamefont
  {L.}~\bibnamefont {Spielberger}}, \bibinfo {author} {\bibfnamefont
  {K.}~\bibnamefont {Khayyat}}, \bibinfo {author} {\bibfnamefont
  {T.}~\bibnamefont {Vogt}}, \bibinfo {author} {\bibfnamefont {H.}~\bibnamefont
  {Br{\"a}uning}}, \bibinfo {author} {\bibfnamefont {K.}~\bibnamefont
  {Ullmann}}, \bibinfo {author} {\bibfnamefont {R.}~\bibnamefont {Moshammer}},
  \bibinfo {author} {\bibfnamefont {J.}~\bibnamefont {Ullrich}}, \bibinfo
  {author} {\bibfnamefont {S.}~\bibnamefont {Hagmann}}, \bibinfo {author}
  {\bibfnamefont {K.-O.}\ \bibnamefont {Groeneveld}}, \bibinfo {author}
  {\bibfnamefont {C.}~\bibnamefont {Cocke}}, \ and\ \bibinfo {author}
  {\bibfnamefont {H.}~\bibnamefont {Schmidt-B{\"o}cking}},\ }\bibfield  {title}
  {\enquote {\bibinfo {title} {Multi-hit detector system for complete momentum
  balance in spectroscopy in molecular fragmentation processes},}\ }\href
  {\doibase http://dx.doi.org/10.1016/S0168-583X(98)00916-1} {\bibfield
  {journal} {\bibinfo  {journal} {Nuclear Instruments and Methods in Physics
  Research Section B: Beam Interactions with Materials and Atoms}\ }\textbf
  {\bibinfo {volume} {149}},\ \bibinfo {pages} {490--500} (\bibinfo {year}
  {1999})}\BibitemShut {NoStop}%
\bibitem [{\citenamefont {Spiegelberg}\ \emph {et~al.}(2017)\citenamefont
  {Spiegelberg}, \citenamefont {Muto}, \citenamefont {Ohtsuka}, \citenamefont
  {Pelckmans},\ and\ \citenamefont {Rusz}}]{SPIEGELBERG2017}%
  \BibitemOpen
  \bibfield  {author} {\bibinfo {author} {\bibfnamefont {J.}~\bibnamefont
  {Spiegelberg}}, \bibinfo {author} {\bibfnamefont {S.}~\bibnamefont {Muto}},
  \bibinfo {author} {\bibfnamefont {M.}~\bibnamefont {Ohtsuka}}, \bibinfo
  {author} {\bibfnamefont {K.}~\bibnamefont {Pelckmans}}, \ and\ \bibinfo
  {author} {\bibfnamefont {J.}~\bibnamefont {Rusz}},\ }\bibfield  {title}
  {\enquote {\bibinfo {title} {Unmixing hyperspectral data by using signal
  subspace sampling},}\ }\href {\doibase
  https://doi.org/10.1016/j.ultramic.2017.07.009} {\bibfield  {journal}
  {\bibinfo  {journal} {Ultramicroscopy}\ }\textbf {\bibinfo {volume} {182}},\
  \bibinfo {pages} {205 -- 211} (\bibinfo {year} {2017})}\BibitemShut {NoStop}%
\bibitem [{\citenamefont {Kruit}, \citenamefont {Shuman},\ and\ \citenamefont
  {Somlyo}(1984)}]{Kruit1984}%
  \BibitemOpen
  \bibfield  {author} {\bibinfo {author} {\bibfnamefont {P.}~\bibnamefont
  {Kruit}}, \bibinfo {author} {\bibfnamefont {H.}~\bibnamefont {Shuman}}, \
  and\ \bibinfo {author} {\bibfnamefont {A.}~\bibnamefont {Somlyo}},\
  }\bibfield  {title} {\enquote {\bibinfo {title} {{Detection of X-rays and
  electron energy loss events in time coincidence}},}\ }\href {\doibase
  10.1016/0304-3991(84)90199-2} {\bibfield  {journal} {\bibinfo  {journal}
  {Ultramicroscopy}\ }\textbf {\bibinfo {volume} {13}},\ \bibinfo {pages}
  {205--213} (\bibinfo {year} {1984})}\BibitemShut {NoStop}%
\bibitem [{\citenamefont {Reimer}\ and\ \citenamefont
  {Kohl}(2008)}]{Reimer2008}%
  \BibitemOpen
  \bibfield  {author} {\bibinfo {author} {\bibfnamefont {L.}~\bibnamefont
  {Reimer}}\ and\ \bibinfo {author} {\bibfnamefont {H.}~\bibnamefont {Kohl}},\
  }\href@noop {} {\emph {\bibinfo {title} {Transmission Electron Microscopy -
  Physics of Image Formation}}},\ \bibinfo {edition} {5th}\ ed.,\ edited by\
  \bibinfo {editor} {\bibfnamefont {W.~T.}\ \bibnamefont {Rhodes}}\ (\bibinfo
  {publisher} {Springer},\ \bibinfo {year} {2008})\BibitemShut {NoStop}%
\bibitem [{\citenamefont {Shuman}, \citenamefont {Chang},\ and\ \citenamefont
  {Somlyo}(1986)}]{Shuman1986}%
  \BibitemOpen
  \bibfield  {author} {\bibinfo {author} {\bibfnamefont {H.}~\bibnamefont
  {Shuman}}, \bibinfo {author} {\bibfnamefont {C.-F.}\ \bibnamefont {Chang}}, \
  and\ \bibinfo {author} {\bibfnamefont {A.}~\bibnamefont {Somlyo}},\
  }\bibfield  {title} {\enquote {\bibinfo {title} {Elemental imaging and
  resolution in energy-filtered conventional electron microscopy},}\ }\href
  {\doibase https://doi.org/10.1016/0304-3991(86)90201-9} {\bibfield  {journal}
  {\bibinfo  {journal} {Ultramicroscopy}\ }\textbf {\bibinfo {volume} {19}},\
  \bibinfo {pages} {121 -- 133} (\bibinfo {year} {1986})}\BibitemShut {NoStop}%
\bibitem [{\citenamefont {Oelsner}\ \emph {et~al.}(2001)\citenamefont
  {Oelsner}, \citenamefont {Schmidt}, \citenamefont {Schicketanz},
  \citenamefont {Klais}, \citenamefont {Sch{\"o}nhense}, \citenamefont
  {Mergel}, \citenamefont {Jagutzki},\ and\ \citenamefont
  {Schmidt-B{\"o}cking}}]{Oelsner2001}%
  \BibitemOpen
  \bibfield  {author} {\bibinfo {author} {\bibfnamefont {A.}~\bibnamefont
  {Oelsner}}, \bibinfo {author} {\bibfnamefont {O.}~\bibnamefont {Schmidt}},
  \bibinfo {author} {\bibfnamefont {M.}~\bibnamefont {Schicketanz}}, \bibinfo
  {author} {\bibfnamefont {M.}~\bibnamefont {Klais}}, \bibinfo {author}
  {\bibfnamefont {G.}~\bibnamefont {Sch{\"o}nhense}}, \bibinfo {author}
  {\bibfnamefont {V.}~\bibnamefont {Mergel}}, \bibinfo {author} {\bibfnamefont
  {O.}~\bibnamefont {Jagutzki}}, \ and\ \bibinfo {author} {\bibfnamefont
  {H.}~\bibnamefont {Schmidt-B{\"o}cking}},\ }\bibfield  {title} {\enquote
  {\bibinfo {title} {Microspectroscopy and imaging using a delay line detector
  in time-of-flight photoemission microscopy},}\ }\href {\doibase
  http://dx.doi.org/10.1063/1.1405781} {\bibfield  {journal} {\bibinfo
  {journal} {Review of Scientific Instruments}\ }\textbf {\bibinfo {volume}
  {72}},\ \bibinfo {pages} {3968--3974} (\bibinfo {year} {2001})}\BibitemShut
  {NoStop}%
\bibitem [{\citenamefont {M{\"u}ller-Caspary}, \citenamefont {Oelsner},\ and\
  \citenamefont {Potapov}(2015)}]{Muller-Caspary2015a}%
  \BibitemOpen
  \bibfield  {author} {\bibinfo {author} {\bibfnamefont {K.}~\bibnamefont
  {M{\"u}ller-Caspary}}, \bibinfo {author} {\bibfnamefont {A.}~\bibnamefont
  {Oelsner}}, \ and\ \bibinfo {author} {\bibfnamefont {P.}~\bibnamefont
  {Potapov}},\ }\bibfield  {title} {\enquote {\bibinfo {title} {Two-dimensional
  strain mapping in semiconductors by nano-beam electron diffraction employing
  a delay-line detector},}\ }\href {\doibase 10.1063/1.4927837} {\bibfield
  {journal} {\bibinfo  {journal} {Applied Physics Letters}\ }\textbf {\bibinfo
  {volume} {107}},\ \bibinfo {eid} {072110} (\bibinfo {year}
  {2015})}\BibitemShut {NoStop}%
\bibitem [{\citenamefont {Ding}\ \emph {et~al.}(2018)\citenamefont {Ding},
  \citenamefont {Jia}, \citenamefont {Nie}, \citenamefont {Weng}, \citenamefont
  {Cao}, \citenamefont {Chen}, \citenamefont {Wu},\ and\ \citenamefont
  {Liu}}]{Lipeng2018}%
  \BibitemOpen
  \bibfield  {author} {\bibinfo {author} {\bibfnamefont {L.}~\bibnamefont
  {Ding}}, \bibinfo {author} {\bibfnamefont {Z.}~\bibnamefont {Jia}}, \bibinfo
  {author} {\bibfnamefont {J.-F.}\ \bibnamefont {Nie}}, \bibinfo {author}
  {\bibfnamefont {Y.}~\bibnamefont {Weng}}, \bibinfo {author} {\bibfnamefont
  {L.}~\bibnamefont {Cao}}, \bibinfo {author} {\bibfnamefont {H.}~\bibnamefont
  {Chen}}, \bibinfo {author} {\bibfnamefont {X.}~\bibnamefont {Wu}}, \ and\
  \bibinfo {author} {\bibfnamefont {Q.}~\bibnamefont {Liu}},\ }\bibfield
  {title} {\enquote {\bibinfo {title} {The structural and compositional
  evolution of precipitates in al-mg-si-cu alloy},}\ }\href {\doibase
  https://doi.org/10.1016/j.actamat.2017.12.036} {\bibfield  {journal}
  {\bibinfo  {journal} {Acta Materialia}\ }\textbf {\bibinfo {volume} {145}},\
  \bibinfo {pages} {437 -- 450} (\bibinfo {year} {2018})}\BibitemShut {NoStop}%
\bibitem [{\citenamefont {Krause}\ and\ \citenamefont
  {Oliver}(1979)}]{Krause1979}%
  \BibitemOpen
  \bibfield  {author} {\bibinfo {author} {\bibfnamefont {M.~O.}\ \bibnamefont
  {Krause}}\ and\ \bibinfo {author} {\bibfnamefont {J.~H.}\ \bibnamefont
  {Oliver}},\ }\bibfield  {title} {\enquote {\bibinfo {title} {Natural widths
  of atomic k and l levels, kα x‐ray lines and several kll auger lines},}\
  }\href {\doibase 10.1063/1.555595} {\bibfield  {journal} {\bibinfo  {journal}
  {Journal of Physical and Chemical Reference Data}\ }\textbf {\bibinfo
  {volume} {8}},\ \bibinfo {pages} {329--338} (\bibinfo {year} {1979})},\
  \Eprint {http://arxiv.org/abs/https://doi.org/10.1063/1.555595}
  {https://doi.org/10.1063/1.555595} \BibitemShut {NoStop}%
\bibitem [{\citenamefont {Voreades}(1976)}]{VOREADES1976}%
  \BibitemOpen
  \bibfield  {author} {\bibinfo {author} {\bibfnamefont {D.}~\bibnamefont
  {Voreades}},\ }\bibfield  {title} {\enquote {\bibinfo {title} {Secondary
  electron emission from thin carbon films},}\ }\href {\doibase
  https://doi.org/10.1016/0039-6028(76)90320-4} {\bibfield  {journal} {\bibinfo
   {journal} {Surface Science}\ }\textbf {\bibinfo {volume} {60}},\ \bibinfo
  {pages} {325 -- 348} (\bibinfo {year} {1976})}\BibitemShut {NoStop}%
\bibitem [{\citenamefont {Mullejans}\ \emph {et~al.}(1993)\citenamefont
  {Mullejans}, \citenamefont {Bleloch}, \citenamefont {Howie},\ and\
  \citenamefont {Tomita}}]{Mullejans1993}%
  \BibitemOpen
  \bibfield  {author} {\bibinfo {author} {\bibfnamefont {H.}~\bibnamefont
  {Mullejans}}, \bibinfo {author} {\bibfnamefont {A.}~\bibnamefont {Bleloch}},
  \bibinfo {author} {\bibfnamefont {A.}~\bibnamefont {Howie}}, \ and\ \bibinfo
  {author} {\bibfnamefont {M.}~\bibnamefont {Tomita}},\ }\bibfield  {title}
  {\enquote {\bibinfo {title} {Secondary electron coincidence detection and
  time of flight spectroscopy},}\ }\href {\doibase
  https://doi.org/10.1016/0304-3991(93)90047-2} {\bibfield  {journal} {\bibinfo
   {journal} {Ultramicroscopy}\ }\textbf {\bibinfo {volume} {52}},\ \bibinfo
  {pages} {360 -- 368} (\bibinfo {year} {1993})}\BibitemShut {NoStop}%
\bibitem [{\citenamefont {Graham}, \citenamefont {Spence},\ and\ \citenamefont
  {Alexander}(1986)}]{Graham1986}%
  \BibitemOpen
  \bibfield  {author} {\bibinfo {author} {\bibfnamefont {R.~J.}\ \bibnamefont
  {Graham}}, \bibinfo {author} {\bibfnamefont {J.}~\bibnamefont {Spence}}, \
  and\ \bibinfo {author} {\bibfnamefont {H.}~\bibnamefont {Alexander}},\
  }\bibfield  {title} {\enquote {\bibinfo {title} {Infrared cathodoluminescence
  studies from dislocations in silicon in tem, a fourier transform spectrometer
  for cl in tem and els/cl coincidence measurements of lifetimes in
  semiconductors},}\ }\href {\doibase 10.1557/PROC-82-235} {\bibfield
  {journal} {\bibinfo  {journal} {MRS Proceedings}\ }\textbf {\bibinfo {volume}
  {82}},\ \bibinfo {pages} {235} (\bibinfo {year} {1986})}\BibitemShut
  {NoStop}%
\bibitem [{\citenamefont {Meuret}\ \emph {et~al.}(2015)\citenamefont {Meuret},
  \citenamefont {Tizei}, \citenamefont {Cazimajou}, \citenamefont
  {Bourrellier}, \citenamefont {Chang}, \citenamefont {Treussart},\ and\
  \citenamefont {Kociak}}]{Meuret2015}%
  \BibitemOpen
  \bibfield  {author} {\bibinfo {author} {\bibfnamefont {S.}~\bibnamefont
  {Meuret}}, \bibinfo {author} {\bibfnamefont {L.~H.~G.}\ \bibnamefont
  {Tizei}}, \bibinfo {author} {\bibfnamefont {T.}~\bibnamefont {Cazimajou}},
  \bibinfo {author} {\bibfnamefont {R.}~\bibnamefont {Bourrellier}}, \bibinfo
  {author} {\bibfnamefont {H.~C.}\ \bibnamefont {Chang}}, \bibinfo {author}
  {\bibfnamefont {F.}~\bibnamefont {Treussart}}, \ and\ \bibinfo {author}
  {\bibfnamefont {M.}~\bibnamefont {Kociak}},\ }\bibfield  {title} {\enquote
  {\bibinfo {title} {Photon bunching in cathodoluminescence},}\ }\href
  {\doibase 10.1103/PhysRevLett.114.197401} {\bibfield  {journal} {\bibinfo
  {journal} {Phys. Rev. Lett.}\ }\textbf {\bibinfo {volume} {114}},\ \bibinfo
  {pages} {197401} (\bibinfo {year} {2015})}\BibitemShut {NoStop}%
\bibitem [{\citenamefont {Kremer}\ \emph {et~al.}(1988)\citenamefont {Kremer},
  \citenamefont {Barnes}, \citenamefont {Chang}, \citenamefont {Evans},
  \citenamefont {Filippone}, \citenamefont {Hahn},\ and\ \citenamefont
  {Mitchell}}]{Kremer1988}%
  \BibitemOpen
  \bibfield  {author} {\bibinfo {author} {\bibfnamefont {R.~M.}\ \bibnamefont
  {Kremer}}, \bibinfo {author} {\bibfnamefont {C.~A.}\ \bibnamefont {Barnes}},
  \bibinfo {author} {\bibfnamefont {K.~H.}\ \bibnamefont {Chang}}, \bibinfo
  {author} {\bibfnamefont {H.~C.}\ \bibnamefont {Evans}}, \bibinfo {author}
  {\bibfnamefont {B.~W.}\ \bibnamefont {Filippone}}, \bibinfo {author}
  {\bibfnamefont {K.~H.}\ \bibnamefont {Hahn}}, \ and\ \bibinfo {author}
  {\bibfnamefont {L.~W.}\ \bibnamefont {Mitchell}},\ }\bibfield  {title}
  {\enquote {\bibinfo {title} {Coincidence measurement of the
  $^{12}$c(\ensuremath{\alpha},\ensuremath{\gamma}${)}^{16}$o cross section at
  low energies},}\ }\href {\doibase 10.1103/PhysRevLett.60.1475} {\bibfield
  {journal} {\bibinfo  {journal} {Phys. Rev. Lett.}\ }\textbf {\bibinfo
  {volume} {60}},\ \bibinfo {pages} {1475--1478} (\bibinfo {year}
  {1988})}\BibitemShut {NoStop}%
\bibitem [{\citenamefont {de~la Pe{\~{n}}a}\ \emph {et~al.}(2018)\citenamefont
  {de~la Pe{\~{n}}a}, \citenamefont {Fauske}, \citenamefont {Burdet},
  \citenamefont {Prestat}, \citenamefont {Jokubauskas}, \citenamefont {Nord},
  \citenamefont {Ostasevicius}, \citenamefont {MacArthur}, \citenamefont
  {Sarahan}, \citenamefont {Johnstone},\ and\ \citenamefont
  {et~al.}}]{Pena2018}%
  \BibitemOpen
  \bibfield  {author} {\bibinfo {author} {\bibfnamefont {F.}~\bibnamefont
  {de~la Pe{\~{n}}a}}, \bibinfo {author} {\bibfnamefont {V.~T.}\ \bibnamefont
  {Fauske}}, \bibinfo {author} {\bibfnamefont {P.}~\bibnamefont {Burdet}},
  \bibinfo {author} {\bibfnamefont {E.}~\bibnamefont {Prestat}}, \bibinfo
  {author} {\bibfnamefont {P.}~\bibnamefont {Jokubauskas}}, \bibinfo {author}
  {\bibfnamefont {M.}~\bibnamefont {Nord}}, \bibinfo {author} {\bibfnamefont
  {T.}~\bibnamefont {Ostasevicius}}, \bibinfo {author} {\bibfnamefont {K.~E.}\
  \bibnamefont {MacArthur}}, \bibinfo {author} {\bibfnamefont {M.}~\bibnamefont
  {Sarahan}}, \bibinfo {author} {\bibfnamefont {D.~N.}\ \bibnamefont
  {Johnstone}}, \ and\ \bibinfo {author} {\bibnamefont {et~al.}},\ }\bibfield
  {title} {\enquote {\bibinfo {title} {hyperspy/hyperspy v1.4.1},}\ }\href
  {\doibase 10.5281/zenodo.1469364} {\  (\bibinfo {year} {2018}),\
  10.5281/zenodo.1469364}\BibitemShut {NoStop}%
\bibitem [{\citenamefont {Hubbell}(2004)}]{Hubbel2004}%
  \BibitemOpen
  \bibfield  {author} {\bibinfo {author} {\bibfnamefont {J.~H.}\ \bibnamefont
  {Hubbell}},\ }\bibfield  {title} {\enquote {\bibinfo {title} {Erratum: ‘a
  review bibliography and tabulation of k l and higher atomic shell x-ray
  fluorescence yields},}\ }\href {\doibase https://doi.org/10.1063/1.1756152}
  {\bibfield  {journal} {\bibinfo  {journal} {J. Phys. Chem. Reference Data}\
  }\textbf {\bibinfo {volume} {33}},\ \bibinfo {pages} {621} (\bibinfo {year}
  {2004})}\BibitemShut {NoStop}%
\end{thebibliography}%

\end{document}